\documentclass[twoside,twocolumn,9pt]{article}
\usepackage{extsizes}
\usepackage[super,sort&compress,comma]{natbib} 
\usepackage[version=3]{mhchem}
\usepackage[left=1.5cm, right=1.5cm, top=1.785cm, bottom=2.0cm]{geometry}
\usepackage{balance}
\usepackage{times,mathptmx}
\usepackage{sectsty}
\usepackage{graphicx} 
\usepackage{lastpage}
\usepackage[format=plain,justification=justified,singlelinecheck=false,font={stretch=1.125,small,sf},labelfont=bf,labelsep=space]{caption}
\usepackage{float}
\usepackage{fancyhdr}
\usepackage{fnpos}
\usepackage[english]{babel}
\usepackage{array}
\usepackage{droidsans}
\usepackage{charter}
\usepackage[T1]{fontenc}
\usepackage[usenames,dvipsnames]{xcolor}
\usepackage{setspace}
\usepackage[compact]{titlesec}
\usepackage{epstopdf}%This line makes .eps figures into .pdf - please comment out if not required.

\definecolor{cream}{RGB}{222,217,201}

\begin{document}

\pagestyle{fancy}
\thispagestyle{plain}
\fancypagestyle{plain}{

%%%HEADER%%%
\fancyhead[C]{\includegraphics[width=18.5cm]{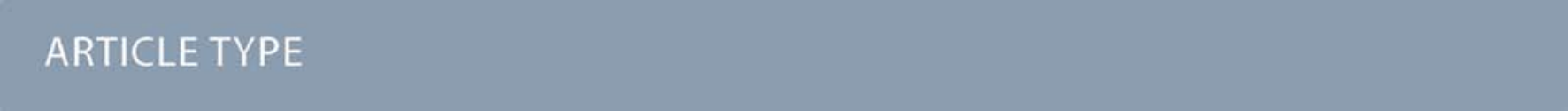}}
\fancyhead[L]{\hspace{0cm}\vspace{1.5cm}\includegraphics[height=30pt]{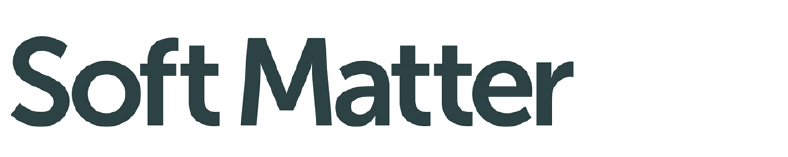}}
\fancyhead[R]{\hspace{0cm}\vspace{1.7cm}\includegraphics[height=55pt]{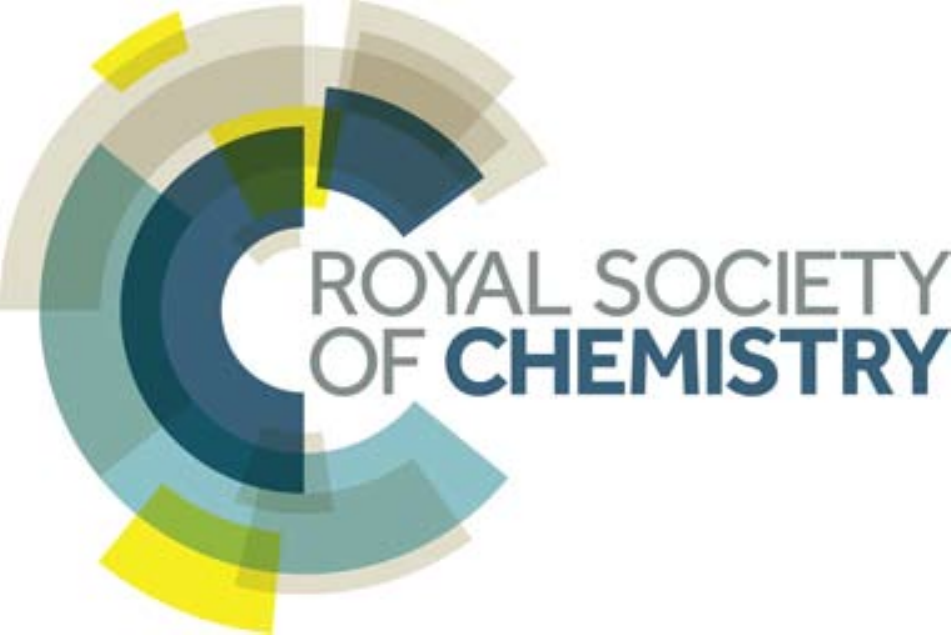}}
\renewcommand{\headrulewidth}{0pt}
}
%%%END OF HEADER%%%

%%%PAGE SETUP - Please do not change any commands within this section%%%
\makeFNbottom
\makeatletter
\renewcommand\LARGE{\@setfontsize\LARGE{15pt}{17}}
\renewcommand\Large{\@setfontsize\Large{12pt}{14}}
\renewcommand\large{\@setfontsize\large{10pt}{12}}
\renewcommand\footnotesize{\@setfontsize\footnotesize{7pt}{10}}
\makeatother

\renewcommand{\thefootnote}{\fnsymbol{footnote}}
\renewcommand\footnoterule{\vspace*{1pt}% 
\color{cream}\hrule width 3.5in height 0.4pt \color{black}\vspace*{5pt}} 
\setcounter{secnumdepth}{5}

\makeatletter 
\renewcommand\@biblabel[1]{#1}            
\renewcommand\@makefntext[1]% 
{\noindent\makebox[0pt][r]{\@thefnmark\,}#1}
\makeatother 
\renewcommand{\figurename}{\small{Fig.}~}
\sectionfont{\sffamily\Large}
\subsectionfont{\normalsize}
\subsubsectionfont{\bf}
\setstretch{1.125} %In particular, please do not alter this line.
\setlength{\skip\footins}{0.8cm}
\setlength{\footnotesep}{0.25cm}
\setlength{\jot}{10pt}
\titlespacing*{\section}{0pt}{4pt}{4pt}
\titlespacing*{\subsection}{0pt}{15pt}{1pt}
%%%END OF PAGE SETUP%%%

%%%FOOTER%%%
\fancyfoot{}
\fancyfoot[LO,RE]{\vspace{-7.1pt}\includegraphics[height=9pt]{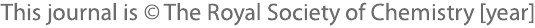}}
\fancyfoot[CO]{\vspace{-7.1pt}\hspace{13.2cm}\includegraphics{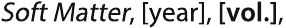}}
\fancyfoot[CE]{\vspace{-7.2pt}\hspace{-14.2cm}\includegraphics{RF}}
\fancyfoot[RO]{\footnotesize{\sffamily{1--\pageref{LastPage} ~\textbar  \hspace{2pt}\thepage}}}
\fancyfoot[LE]{\footnotesize{\sffamily{\thepage~\textbar\hspace{3.45cm} 1--\pageref{LastPage}}}}
\fancyhead{}
\renewcommand{\headrulewidth}{0pt} 
\renewcommand{\footrulewidth}{0pt}
\setlength{\arrayrulewidth}{1pt}
\setlength{\columnsep}{6.5mm}
\setlength\bibsep{1pt}
%%%END OF FOOTER%%%

%%%FIGURE SETUP - please do not change any commands within this section%%%
\makeatletter 
\newlength{\figrulesep} 
\setlength{\figrulesep}{0.5\textfloatsep} 

\newcommand{\topfigrule}{\vspace*{-1pt}% 
\noindent{\color{cream}\rule[-\figrulesep]{\columnwidth}{1.5pt}} }

\newcommand{\botfigrule}{\vspace*{-2pt}% 
\noindent{\color{cream}\rule[\figrulesep]{\columnwidth}{1.5pt}} }

\newcommand{\dblfigrule}{\vspace*{-1pt}% 
\noindent{\color{cream}\rule[-\figrulesep]{\textwidth}{1.5pt}} }

\makeatother
%%%END OF FIGURE SETUP%%%

%%%TITLE, AUTHORS AND ABSTRACT%%%
\twocolumn[
  \begin{@twocolumnfalse}
\vspace{3cm}
\sffamily
\begin{tabular}{m{4.5cm} p{13.5cm} }

\includegraphics{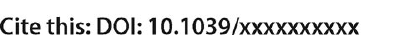} & \noindent\LARGE{\textbf{Splitting droplet through coalescence of two different three-phase contact lines$^\dag$}} \\%Article title goes here instead of the text "This is the title"
\vspace{0.3cm} & \vspace{0.3cm} \\

 & \noindent\large{Haitao Yu, \textit{$^{a\ddag}$}, Pallav Kant,\textit{$^{b\ddag}$}, Brendan Dyett,\textit{$^{a}$}, Detlef Lohse,\textit{$^{b,c}$}, and Xuehua Zhang\textit{$^{d}$}$^{\ast}$} \\%Author names go here instead of "Full name", etc.

\includegraphics{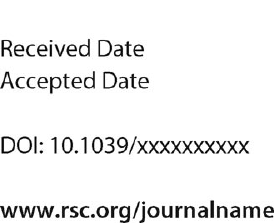} & \noindent\normalsize{Moving contact lines of more than two phases dictate a large number of interfacial phenomena.
Despite of its significance to fundamental and applied processes, the contact lines at a junction of four-phases (two immiscible liquids, solid and gas) have been addressed only in a few investigations.
Here, we report an intriguing phenomenon that follows after the four phases of oil, water, solid and gas make contact through the coalescence of two different three-phase contact lines. 
We combine experimental study and theoretical analysis to reveal and rationalize the dynamics exhibited upon the coalescence  between the contact line of a micron-sized oil droplet and the receding contact line of a millimeter-sized water drop that covers the oil droplet on the substrate.
We find that after the coalescence a four-phase contact line is formed for a brief period. However this quadruple contact line is not stable, leading to a `droplet splitting' effect and eventually expulsion of the oil droplet from the water drop. We then show that the interfacial tension between the different phases and the viscosity of oil droplet dictate the splitting dynamics.
More viscous oils display higher resistance to the extreme deformations of the droplet induced by the instability of quadruple contact line and no droplet expulsion is observed for such cases.
} \\%The abstrast goes here instead of the text "The abstract should be..."

\end{tabular}

 \end{@twocolumnfalse} \vspace{0.6cm}

  ]
%%%END OF TITLE, AUTHORS AND ABSTRACT%%%

%%%FONT SETUP - please do not change any commands within this section
\renewcommand*\rmdefault{bch}\normalfont\upshape
\rmfamily
\section*{}
\vspace{-1cm}

%%%FOOTNOTES%%%

\footnotetext{\textit{$^{a}$~Soft Matter \& Interfaces Group, School of Engineering, Royal Melbourne Institute of Technology University, Melbourne, VIC 3001, Australia.}}
\footnotetext{\textit{$^{b}$~Physics of Fluids Group University of Twente, 7500 AE Enschede, The Netherlands; Max Planck Center Twente for Complex Fluid Dynamics. }}
\footnotetext{\textit{$^{c}$~Max Planck Institute for Dynamics and Self-Organization, 37077 Gottingen, Germany. }}
\footnotetext{\textit{$^{d}$~Department of Chemical and Materials Engineering, University of Alberta, Edmonton, AB T6G 1H9, Canada.}}

%Please use \dag to cite the ESI in the main text of the article.
%If you article does not have ESI please remove the the \dag symbol from the title and the footnotetext below.
\footnotetext{\dag~Electronic Supplementary Information (ESI) available: 
Supporting Video 1 shows the splitting of an octanol droplet with a diameter of  100~$\mu m$ recorded by a bottom-view confocal microscope.  
Supporting Video 2 shows the splitting of an octanol droplet with a diameter of  99~$\mu m$ recorded by synchronized bottom-view and side-view high-speed cameras at 3000 fps.
Supporting Video 3 shows the splitting of a decanol droplet with a diameter of 97~$\mu m$ recorded by a bottom-view high-speed camera at 3000 fps.
See DOI: 10.1039/cXsm00000x/}
%additional addresses can be cited as above using the lower-case letters, c, d, e... If all authors are from the same address, no letter is required

\footnotetext{\ddag~H.Y. and P.K. contributed equally to this work.}
\footnotetext{$^{\ast}$~To whom correspondence should be addressed: xuehua.zhang@ualberta.ca}

%%%END OF FOOTNOTES%%%

%%%MAIN TEXT%%%%
\section{Introduction}

The dynamics of contact lines (CLs) formed at the contact of a solid, a liquid and a gaseous phase is a topic of long-standing interest as such triple lines are commonly encountered in our daily lives and in numerous industrial processes \cite{bonn,deGennes}. 
However, in view of recent innovations in the fields of inkjet printing based manufacturing \cite{sun2015droplet}, material processing \cite{sun2003compound}, and micro/nanofluidics \cite{utada2005monodisperse,lohse2015}, it has become imperative to develop a detailed understanding of the contact lines formed when more than three phases make contact.
Particularly, contact lines formed at the junction of two immiscible liquids, solid and ambient gaseous phases are often encountered in the scenarios such as spreading of a core-shell droplet on a solid surface \cite{johnson1985fluid,gao2011spreading}, wetting/dewetting of thin-films of binary mixtures \cite{pototsky, nanoStructures}, spreading of a droplet over nanostructured surfaces partially infused with lubricant \cite{wong2011bioinspired, hejazi, varanasi, keiser2017drop, sadullah2018drop}. In a large industrial process of warm water extraction, the dynamics of the contact line formed by oil, water, sand and aerated bubble is crucial for the detachment and efficient separation of heavy oils from sand grains \cite{lin2014dewetting}.

Despite of its significance to fundamental and applied processes, the dynamics of a four phases contact line has been addressed only in a few investigations. 
The theoretical investigations by Mahadevan \textsl{et al} \cite{maha} was first such study in which the statics of a contact line formed at the contact of four different phases was addressed.
In the context of a sessile compound droplet, the authors derived the necessary conditions in terms of the volume ratios and interfacial tensions for the existence of a quadruple contact line at a junction of two immiscible liquids-vapour-solid phases. 
Later, Hejazi and Nosonovsky \cite{hejazi} investigated the wetting transition of an oil droplet placed on a micro-structured surface in a liquid environment.
They found that trapped air pockets on a micro-structured surface give rise to scenarios where four phases can coexist at a contact line, which significantly alters the equilibrium contact angle of an oil droplet.
However, the \emph{dynamics} exhibited by the oil droplet when the four phases make contact at the four-phase contact line was not addressed.
More recently, Weyer \textsl{et al} \cite{weyer} experimentally achieved a stable quadruple contact line for a compound droplet (a small oil droplet encapsulated in a water droplet) on a fibre.
Again, the scope of the investigations by Weyer \textsl{et al} \cite{weyer} was limited to the identification of the equilibrium shapes of such compound droplets.

In contrast to the above studies, here we investigate the dynamical process at the coalescence of two different three-phase contact lines that are formed by  an oil droplet sitting inside a sessile water drop, i.e. oil-water-solid CL, and formed by this sessile drop by gas, water and solid, i.e. air-water-solid CL. Here we refer the small oil phase as the {\it droplet}, while the macroscopic water is referred to as the {\it drop}.
We employ confocal microscopy and high-speed imaging technique to visualise the dynamical process before and after the slowly receding air-water-solid CL touches the oil-water-solid CL. 
Further, we illustrate that the overall phenomenon is strongly governed by the interfacial tensions between the fluid phases and the viscosity of the encapsulated oil droplet.
The findings reported here may have strong bearings on the industrial processes mentioned before.

\section{Results and Discussion}

\begin{figure}
\centering
\includegraphics[clip, trim=0cm 0cm 0cm 0cm, width=0.5\textwidth]{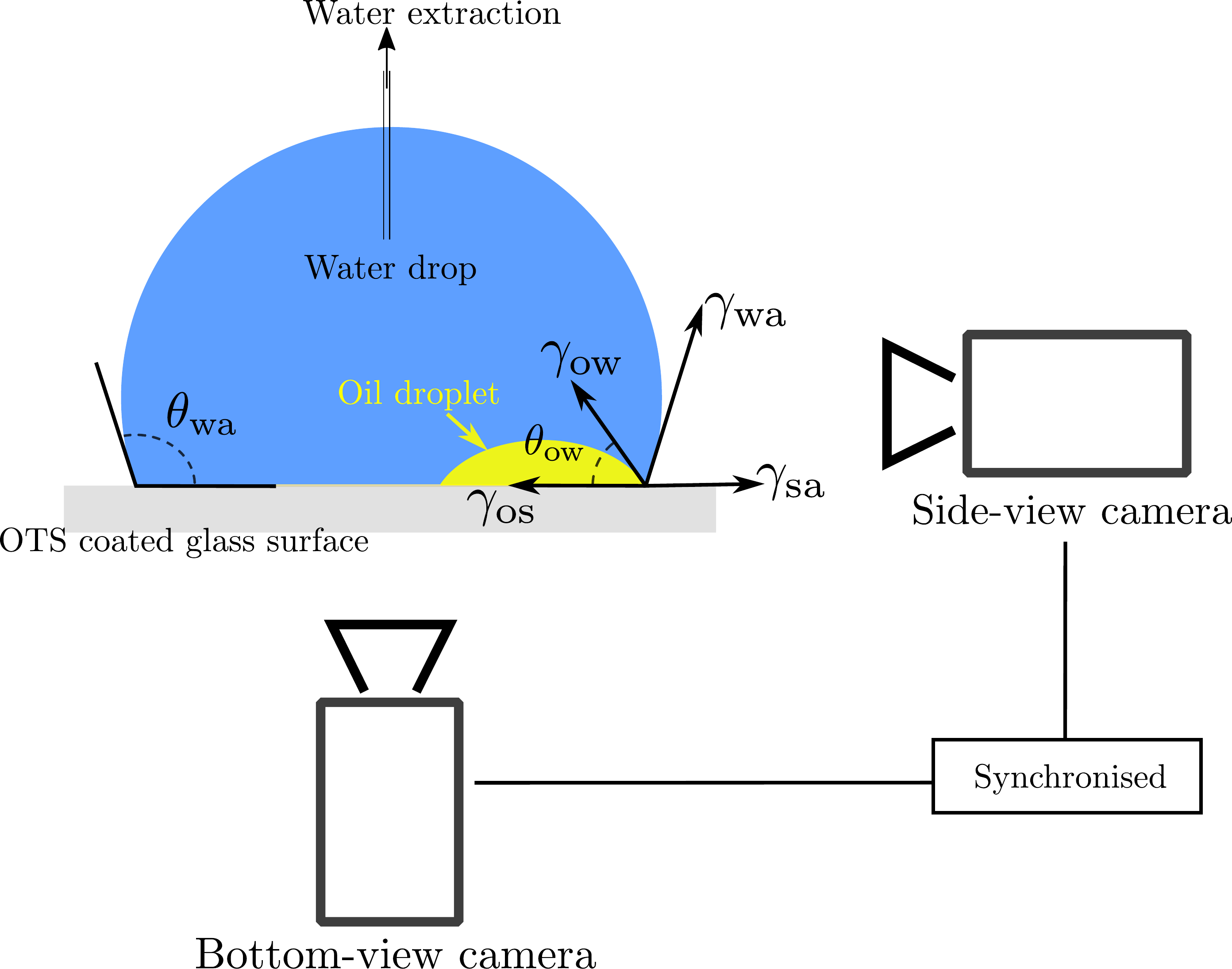}\\
\caption{Schematic drawing of experimental setup. The three-phase contact lines are at the boundaries of the oil droplet (yellow) and of the water drop (blue).  The water was sucked out very slowly until the air-water-solid contact line contacts the encapsulated oil droplet. The dynamical process was recorded using high-speed cameras both in bottom and side view.  Here $\theta_{\mathrm{wa}}$ is the contact angle of a sessile water drop on the hydrophobic substrate in air,  $\theta_{ow}$, the contact angle of an oil droplet in water.
$\gamma_{\mathrm{ow}}$, $\gamma_{\mathrm{os}}$,  $\gamma_{\mathrm{wa}}$, $\gamma_{\mathrm{oa}}$ and  $\gamma_{\mathrm{sa}}$ are the interfacial tensions of two of oil (o), water (w), air (a) or solid (s) phases. For all the oils $\gamma_{\mathrm{oa}} > \gamma_{\mathrm{ow}}$ (Table \ref{tab:table-1}). 
}
\label{fig:figure-1}
\end{figure}

\begin{figure*}
\centering
\includegraphics[clip, trim=0cm 0cm 0cm 0cm, width=0.9\textwidth]{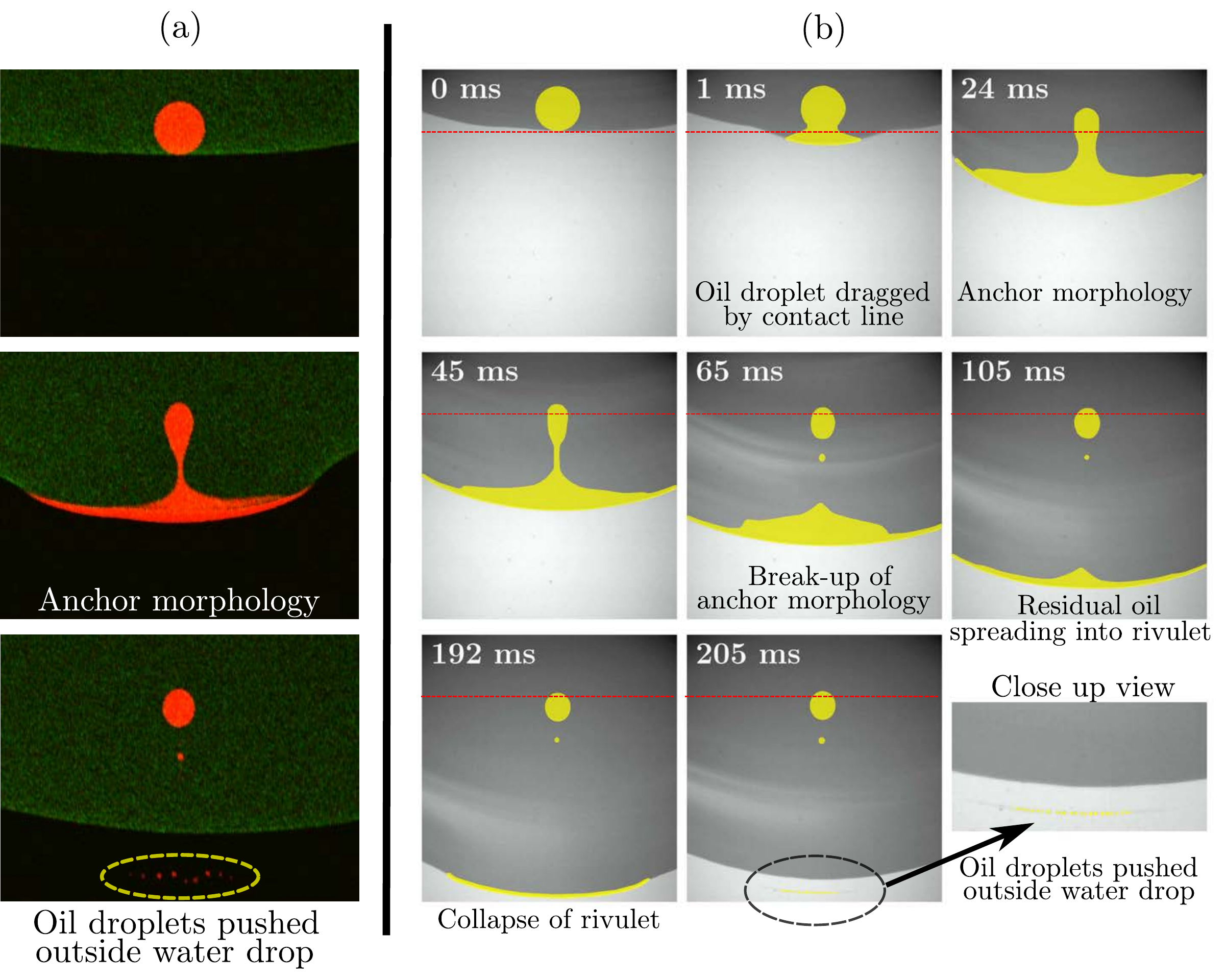}\\
\caption{Sequence of images showing the events after oil, water, air and solid phases make contact. Confocal microscopic images in (a) are taken at 30 fps. Water is dyed green and the oil is blue. High speed optical images in (b) were taken at 3000 fps. The horizontal dotted line on the snapshots indicates the initial position of the air-water-solid CL. The octanol droplet in (b) is highlighted in yellow for better visualisation.}
\label{fig:figure-2}
\end{figure*}

\begin{figure*}
\centering
\includegraphics[width=0.9\textwidth]{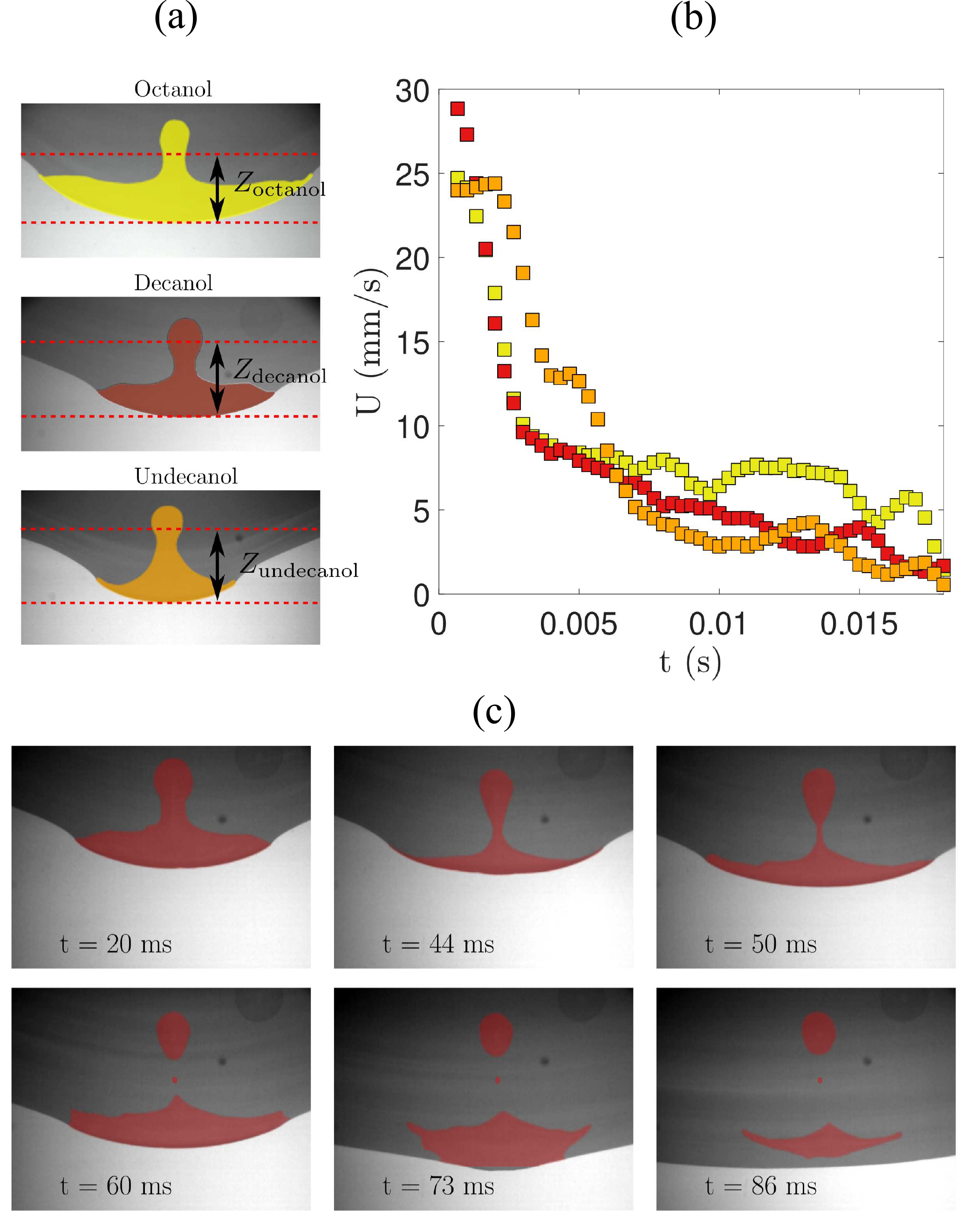}\\
\caption{Effect of the oil viscosity. The snapshots in (a) show the state of deformed contact line and stretched oil droplets of octanol, decanol and undecanol, respectively.  The time t  is 16 ms after the coalescence of the air-water-solid CL and the oil-water-solid CL. The initial contact diameter of all three oil droplets were $\sim~95~\mu$m. The two horizontal red lines indicates the distance that the contact line travelled post the coalescence event. (b) Plot of the velocity of the contact line versus time, measured for three oil droplets of same size. 
(c) Sequence of snapshots showing the separation of decanol and water phases at the advanced four-phase contact line. Yellow--octanol, red--decanol, and orange--undecanol. 
}
\label{fig:figure-4}
\end{figure*}

In the experiments, a droplet of octanol (oil phase) was deposited onto a hydrophobilized glass surface. 
Subsequently a water drop was gently deposited on the substrate, covering the oil droplet. The footprint diameter of the oil droplet and the water drop was 50-500 $\mu m$ and 3-4 $mm$, respectively.
Water was then slowly withdrawn from the top to imparted a slow receding velocity of air-water-solid CL.
The video recordings were triggered when the receding contact line meets oil-water-solid CL.

The sequences of snapshots in Figure \ref{fig:figure-2} and Supporting Videos 1 and 2 reveal a typical series of events. 
At first, as air-water-solid CL approaches the oil droplet, the water slowly drains out from the gap between the two contact lines.
Immediately ($t = 0$ s) upon the coalescence of these two contact lines, a four-phase contact line is formed. 
However, this four-phase contact line is unstable. In less than 1 ms, the drop periphery instantly bulges out towards the gas phase and begins to move in the direction opposite to the externally imposed receding motion.

Interestingly, the moving contact line pulls along the oil droplet, stretching the oil droplet from its initial position. 
Simultaneously, the oil also spreads further along the periphery of the water drop, conforming the initial circular boundary of the oil droplet into an anchor shape ($t = 0.024$ s). With time the moving contact line leads to the elongation of the anchor morphology.
Consequently, the filament connecting the oil droplet and the crown of the anchor becomes very narrow ($t = 0.045$ s). 
Eventually, this filament becomes unstable and breaks up into smaller droplets \textsl{via} a Rayleigh-Plateau type instability \cite{davis, sonin} ($t = 0.065$ s).
The residual droplets remain encapsulated within the water drop and quickly assume a spherical-cap shape with nearly circular footprint due to interfacial tension. 
After this splitting of the oil droplet, the oil fraction carried by the moving contact line continues spreading into a thinning rivulet along the drop periphery ($t = 0.105$ s).

At a turning point with the velocity $U$ of zero, the moving contact line reverses the direction, leaving behind the thin rivulet of oil on the dry surface. 
This oil rivulet soon breaks into multiple microdroplets to minimise the interfacial energy. With time the receding air-water-solid CL reaches the oil-water-solid CL of the residue oil droplet. The same process of droplet splitting repeats till all the oil is delivered outside the water drop. 

The above described processes are captured in supplementary movie M1. The distance travelled by the tip of the moving contact line is measured as a function of time. The data plotted in Figure \ref{fig:figure-4} shows that the advancing velocity  $U$ starts from the maximum $U_{\mathrm{max}}$ immediately at the coalescence event and decreases rapidly within 5 ms and then gradually decays to zero after approximately 15 ms.

The general features of the above droplet splitting process are universal, confirmed in our experiments for various sizes of encapsulated oil droplets with an initial diameter 50--500~$\mu$m and for different imposed receding velocity of air-water-solid CL from 3 to 70~$\mu$m/s.   
Qualitatively, no change was observed as a result of either the difference in the size of the oil droplet or in receding velocity of the air-water-solid CL. Subtle difference was noticed in the experiments performed at higher water withdrawal rates, that is, a slight reduction was noted in the values of $U_{\mathrm{max}}$ and the distance  $Z$ traversed by the tip of the contact line before the turning point. This reduction is presumably due to the larger negative pressure imposed to reach faster receding rates.

In addition to octanol, the experiments were also performed with undecanol and decanol droplets shown in Supporting Video 3.   
 As listed in Table \ref{tab:table-1}, the viscosities of decanol and undecanol are significantly larger than octanol while the surface tensions of all three oils are similar. Figure \ref{fig:figure-4}a show the morphologies of the three oil droplets ($\sim 95~\mu m$ in diameter) at $t=16$ ms after the coalescence of the two three-phase CLs. 
It can be seen from Figure \ref{fig:figure-4}b that the distance travelled by the tip of the moving contact line $Z$ is almost the same for all three oil droplets. 
Moreover, the temporal variation of the moving velocity measured is alike. 
The results imply that the viscosity of the oil droplet has a negligible effect on the instability of the four-phase contact line and the early stage dynamics of the moving contact line.

In a longer term, more viscous oil droplets can be also split into two parts: an oil rivulet along the advancing four-phase CL and a smaller spherical cap droplet at the initial position.  However, the distinct difference is that the least viscous oil (octanol) spreads to a much larger area in comparison to the other two more viscous oils.  
Correspondingly, the localized deformation of the water drop boundary is maximum in the case of the least viscous octanol droplet.
This result suggests a strong influence of the oil viscosity on the long-term dynamics of the moving contact line.
Even more, as the perturbed CL continues to advance, the rivulet of the viscous oil detaches from the moving front, being stranded inside the water drop as shown in the Figure \ref{fig:figure-4}c.  The air-water-solid CL keeps advancing for a longer distance before the turning point, then reverses the direction and retracts. Such separation of the oil rivulet has never been observed for the least viscous oil (octanol) droplet, suggesting a direct consequence of the strong viscous dissipation within the thin oil film that are stretched and dragged by the moving contact line.

More quantitative insight into the dynamical proces described above can be gained from the following theoretical analysis. For a three-phase contact line, the equilibrium condition in both horizontal and vertical direction can be written in terms of the interfacial tensions between two phases, as described by Young \cite{young}. We extend the same formalism to our case when four phases meet at a contact line. 
Similar approach was also adopted by Mahadevan \textsl {et al} \cite{maha} for a static four-phase contact line.
We assume that the water drop and the oil droplet are in a state of quasi-equilibrium at the moment when the two three phase contact lines meet.
Since in our case the solid phase is undeformable, only the force balance in the horizontal direction is of importance, which can be written as:
\begin{equation}
\begin{split}
F_{\mathrm{horizontal}} &=  \gamma_{\mathrm{wa}} cos (\pi - \theta_{\mathrm{wa}}) + \gamma_{\mathrm{sa}} - \gamma_{\mathrm{ow}} cos \theta_{\mathrm{ow}} -\gamma_{\mathrm{os}}, \\
 & = \gamma_{\mathrm{wa}} cos (\pi - \theta_{\mathrm{wa}}) + \gamma_{\mathrm{oa}} cos \theta_{\mathrm{oa}} - \gamma_{ow} cos \theta_{ow},
\label{eq:eqn2}
\end{split}
\end{equation}
where $\theta_{\mathrm{wa}}$ and $\theta_{\mathrm{oa}}$ are the contact angles of a sessile water droplet and oil droplet on our solid surface in dry ambient conditions, respectively.  $\theta_{ow}$ is the contact angle of an oil droplet in wet ambient conditions as sketched in Figure \ref{fig:figure-1}.
Since at the time of coalescence of the two contact lines, the air-water-solid contact line is receding, therefore $\theta_{\mathrm{wa}}$ is the receding contact angle of water ($\sim 100^\circ$). 
Our contact angle measurements yielded $\theta_{\mathrm{oa}} \sim 30^\circ$ and $\theta_{ow} \sim35^\circ$ for all three oils.  Moreover, the interfacial tension of oil and air $\gamma_{\mathrm{oa}}$ is larger than that of oil and water $\gamma_{\mathrm{ow}}$ (Table \ref{tab:table-1}).  Therefore, in our system a non zero horizontal force, $F_{\mathrm{horizontal}} > 0$, acts at the contact line when the four phases make contact.

This horizontal force instantly pushes the contact line outwards to spread the drop onto the dry surface, in the direction opposite to the externally imposed receding motion.
Thus accordingly the contact line locally deforms (bulges towards the gaseous phase), as observed in the experiments.
Most importantly, this also means that in our system the quadruple contact line formed can never attain an equilibrium state.

\begin{figure}
\centering
\includegraphics[clip, trim=0cm 0cm 0cm 0cm, width=0.5\textwidth]{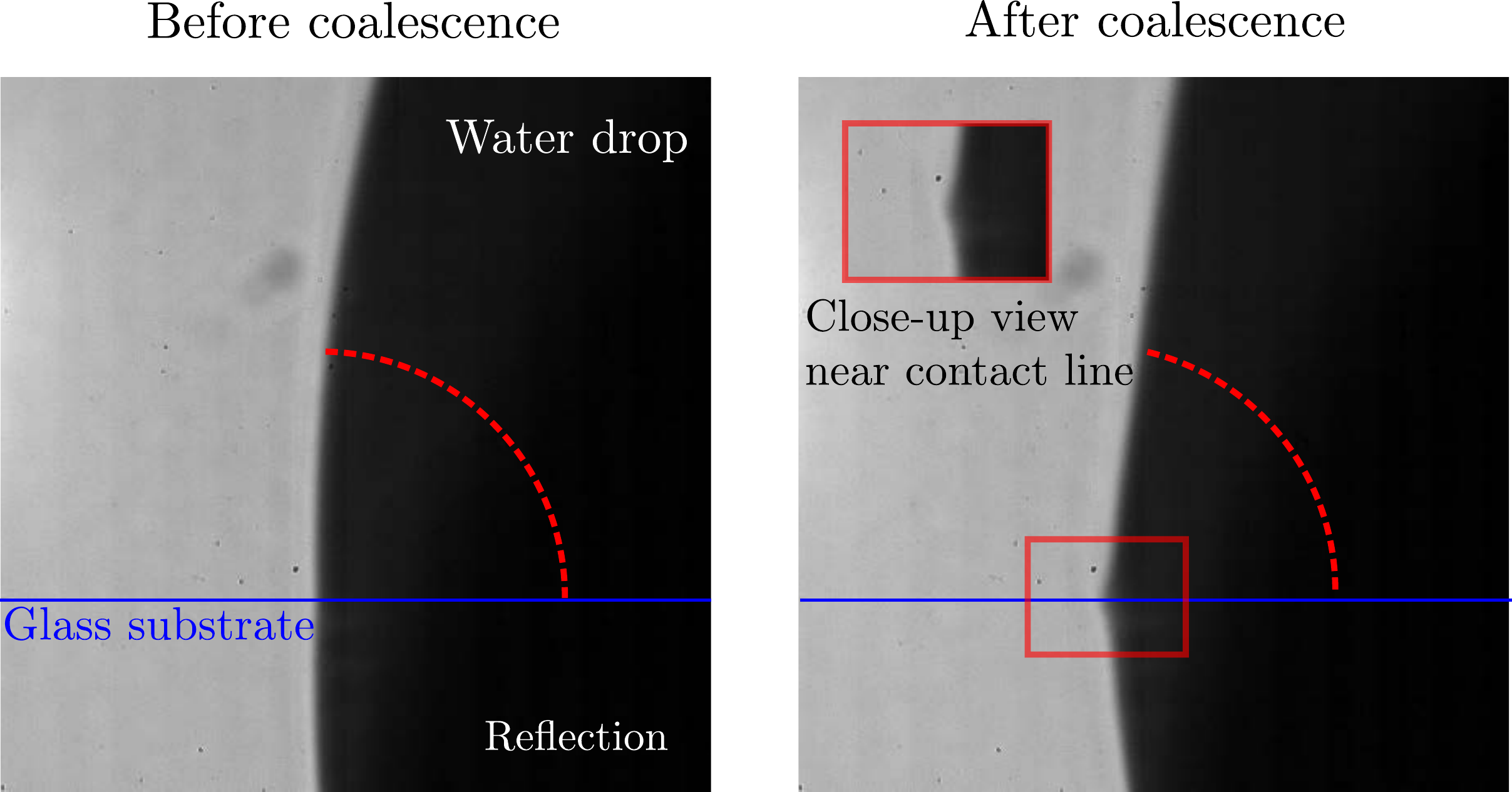}\\
\caption{Immediate change in the contact angle after the merger of two different contact lines. The contact angle sharply reduced to a lower value of 70$^\circ$ after the coalescence. Also note the apparent discontinuity in the slope of the interface near the contact line.}
\label{fig:CA-snapshot}
\end{figure}

\begin{figure*}
\centering
\includegraphics[clip, trim=0cm 0cm 0cm 0cm, width=0.7\textwidth]{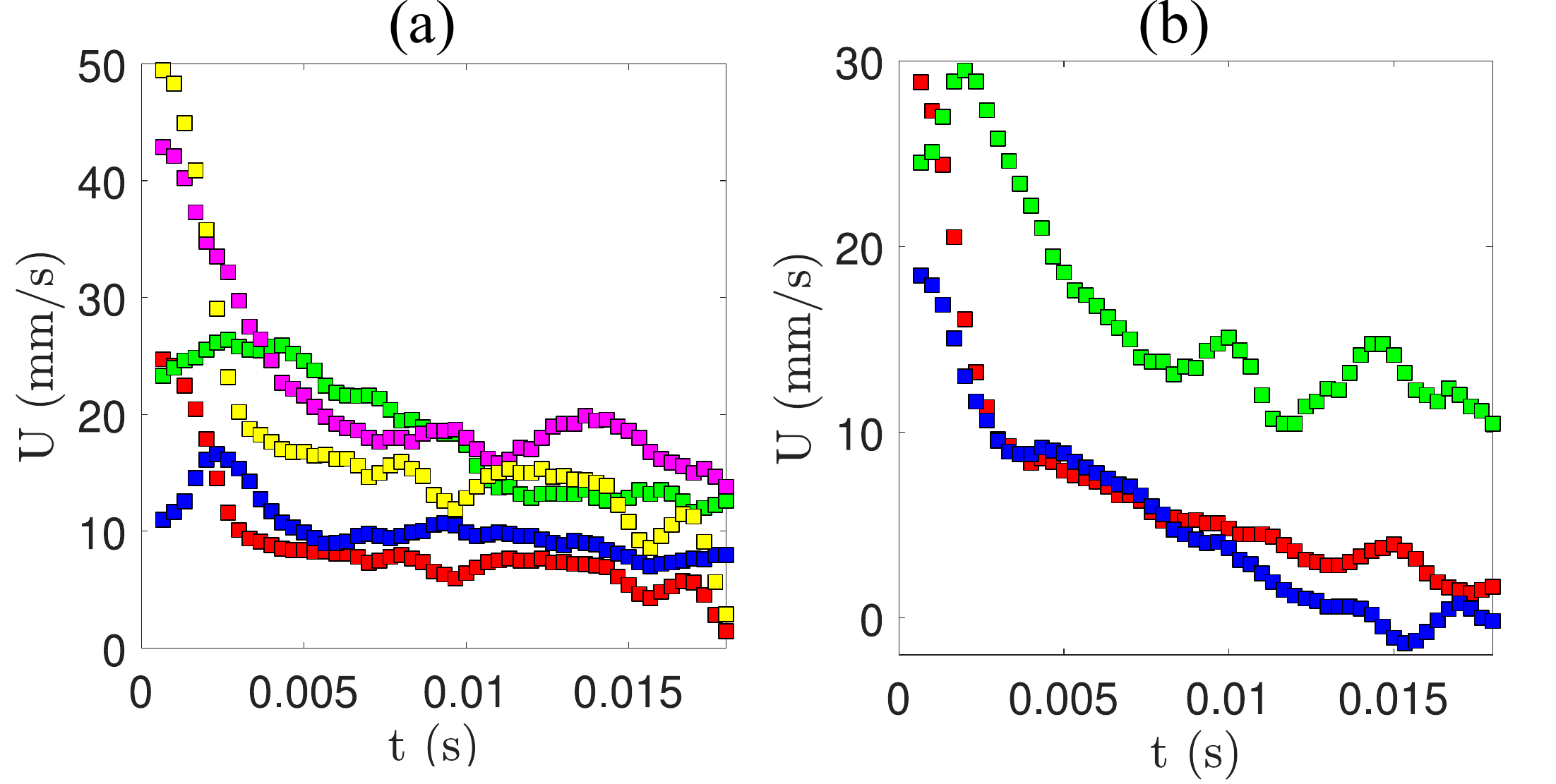}\\
\caption{Temporal variation in the four-phase contact line velocity. The velocity was measure for (a) octanol droplets of different sizes; different colors in the graph indicate the measurements for oil droplets of different footprint diameter, red--99~$\mu$m, green--165~$\mu$m, magenta--342~$\mu$m, yellow--425~$\mu$m and blue--525~$\mu$m. (b) Decanol droplets, red--68~$\mu$m, blue--88~$\mu$m, green--291~$\mu$m.}
\label{fig:CA}
\end{figure*}

Further our side-view measurements reveal that the dynamic contact angle of the moving contact line before and after the coalescence event changes significantly.
It is seen that the moments before the coalescence of the two three-phase CLs, the angle subtended by the water drop is in the range  90$^\circ$--100$^\circ$, the receding contact angle of water on the substrate.
Immediately after the coalescence, the contact angle at the moving contact line reduces to $\sim 70^\circ$  as highlighted in Figure \ref{fig:CA-snapshot}. Importantly,  this angle of $\sim 70^\circ$ is lower than the advancing angle of water drop $\theta_{\mathrm{aw}} \sim 110^\circ$, but higher than that of an oil drop on the same dry substrate $\theta_{\mathrm{oa}} \sim 35^\circ$ for all three oils.

To understand the sequence of events immediately after the coalescence of the two three-phase contact lines, we define the spreading coefficient, $S = \gamma_{\mathrm{wg}} - \gamma_{\mathrm{ow}} -\gamma_{\mathrm{oa}}$.
We find that for all the three employed oils, $S>0$. This suggests that it is energetically favourable for the oil to move outside the water drop and spread over the air-water interface. Similar spontaneous spreading of oil over the interface of a water droplet has also been reported in ref.\cite{carlson, varanasi}. 
This initial spreading of oil over the air-water interface creates a difference in the local interfacial tension over the water drop. The interfacial tension is higher for clean water surface, but lower in the area coated with the oil. Such difference in interfacial intension creates a Marangoni stress, leading to the continuous spreading of oil over the air-water interface as in a perfectly wetting scenario. In our experiments, this spontaneous spreading of oil over the air-water interface provides the driving force that continuously pulls the oil from the encapsulated droplet towards the boundary sliding to the dry surface.

\begin{figure}
\centering
\includegraphics[clip, trim=0cm 0cm 0cm 0cm, width=0.5\textwidth]{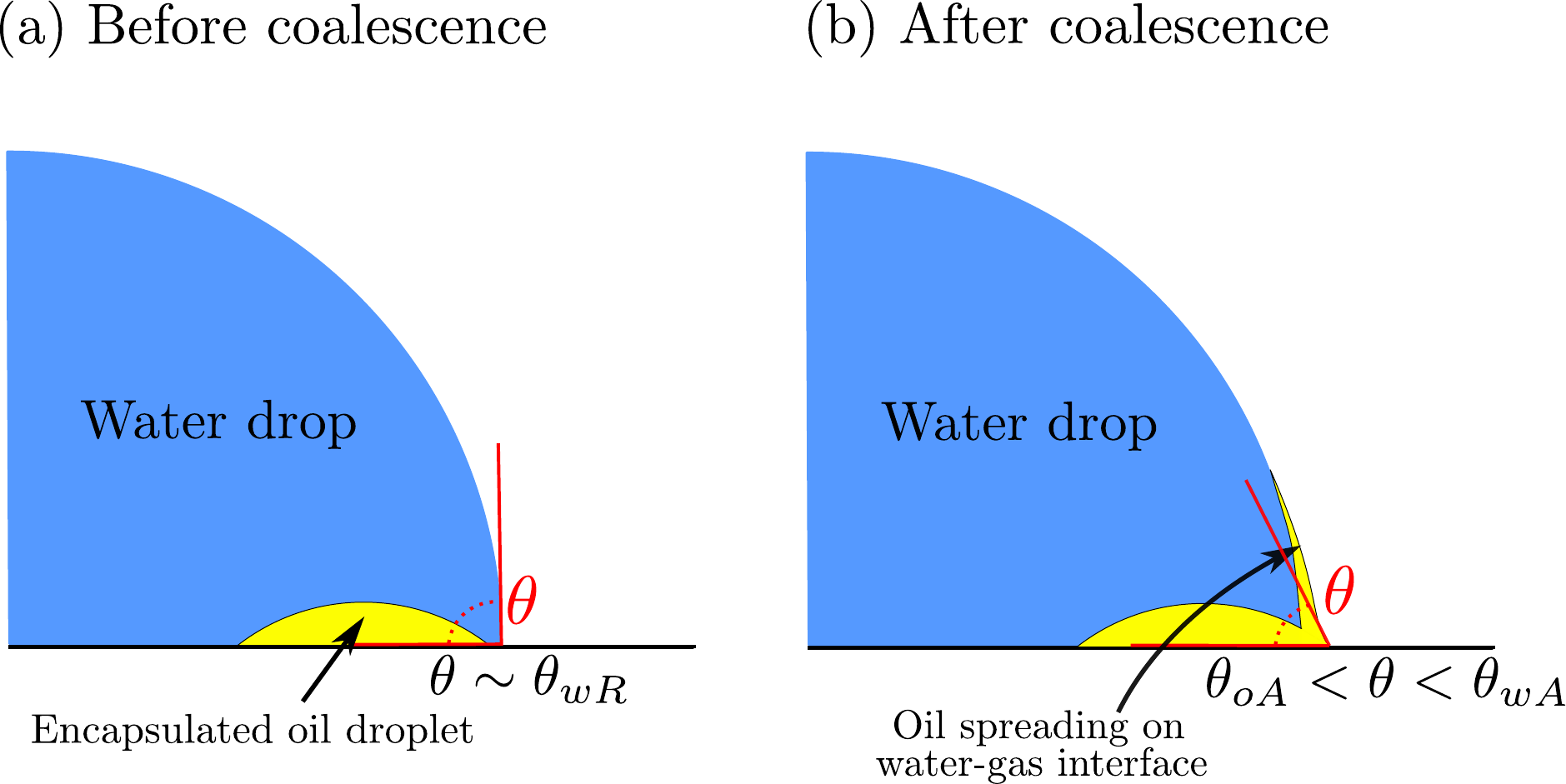}\\
\caption{Schematic diagram showing the arrangement of air, oil, water and solid before and after the coalescence of the air-water-solid and oil-water-solid CLs.  Post the coalescence, the oil pushes across the air-water interface.  The advancing four-phase contact line is formed at the junction of air, oil, water and solid phases. %The dynamic contact angle associated with the advancing contact line after the coalescence is greater(smaller) than the advancing contact angle of an oil(water) droplet on the OTS coated glass surface, $\theta_{\mathrm{oA}} < \theta < \theta_{\mathrm{wA}}$.
}
\label{fig:figure-6}
\end{figure}

When the oil pushes across the air-water interface, the configuration of the four phases may be represented by the schematic diagram in Figure \ref{fig:figure-6}.  Such configuration suggests that the interfacial properties of air, oil and solid govern both the angle and velocity of the moving contact line, as water is insulated by the oil from both solid and air phases. %advancing motion of the contact line at contact angles lower than 110$^\circ$ is only possible if the contact line is formed at the junction of oil-gas-solid phases. 
We confirmed this conjecture by comparing the experimental measured advancing velocities of contact lines formed at the junctions of different oils-gas-solid phases with the theoretical predictions.

We use Cox-Voinov spreading law \cite{cox,voinov} to theoretically compute the velocity of a contact line of air, oil and solid advancing with a contact angle of $70^o$, 
\begin{equation}
U = \gamma_{oa}~(\theta^{3} - \theta_{\mathrm{oa}}^{3})/ (9~\mu~\mathrm{ln}(R/L)),
\end{equation}
where, $\theta_{\mathrm{oa}}$ is the advancing contact angle of an air-oil interface on the substrate, $\simeq 40^\circ$, $R$ is the radius of the drop, same as that of encapsulated oil droplet.  $L$ is the microscopic cut-off length, $\sim 1 A^\circ$, $\mu$ is the viscosity of the oil and $\gamma_{oa}$ is the interfacial tension between oil and gas. 
The computed velocities are shown in the table \ref{tab:table-1}.
Remarkably these theoretical values are of the same order of magnitude as the characteristic velocities, $U_{\mathrm{max}}$, measured in the experiments for octanol and decanol droplets with different initial sizes; see Figure \ref{fig:CA}.
The good agreement between the theoretical and experimental values of advancing velocities supports the above-described mechanism that upon the contact of the four phases, the encapsulated oil moves outside the water drop to spread over the air-water interface and the advancing contact line is formed at the junction of air-water-soild phases.

\section{Conclusion}

In summary, we investigated the dynamics exhibited upon the coalescence of the two different contact lines of an encapsulated oil droplet with that of a large water droplet covering it.
We find that for a brief moment, the contact of the four phases leads to formation of a quadruple contact line.
However, using a formalism similar to what leads to the Young equation \cite{young}, we showed that in our case the quadruple contact line can never attain a state of static equilibrium.
In addition, this formulation also explains the localized outward motion of the contact line observed in the experiments after the contact of the four phases.
Since for the different oils used in the experiments the spreading coefficient $S\,= \gamma_{\mathrm{wg}} - \gamma_{\mathrm{ow}} -\gamma_{\mathrm{oa}}\, >\,0$, the encapsulated oil moves outside the water droplet to spread over the water-vapour interface.
This spreading over the water-vapour interface as in the perfectly wetting scenario is driven by the Marangoni like stresses.
Hence, as a combined effect of the outward motion of the contact line and the Marangoni forcing at the water-vapor interface, the oil droplet is dragged in the direction of motion.
This leads to the fragmentation of the initial droplet via a Rayleigh-Plateau type instability.
We also demonstrated that the viscosity of the oil plays an important role in determining the overall dynamics after the contact of the four phases.
For more viscous oils the outward moving contact line fails to drag droplets along with it due to the stronger viscous dissipation within the thin oil filament.
This leads to the separation of the oil and water phases at the contact line.

\section{Materials and Methods}

	A schematic diagram of the experimental setup is shown in Figure \ref{fig:figure-1}.
The oil droplet was deposited on the hydrophobic substrate by carefully touching a thin metallic wire covered with the oil to the glass surface. The footprint diameter of the oil droplet was 50--500 $\mu$m.
The oils used for experiments were 1-Octanol (Sigma, 99\%) 1-Decanol (Sigma, 98\%), 1-Undecanol (Sigma, 99\%).
The physical properties of the oils used are tabulated in table~\ref{tab:table-1}.
The hydrophobic substrates were  OTS (Octa-decyl-trichlorosiliane) coated glass surface, prepared by following the protocol reported in Zhang \textsl{et al}. \cite{OTS}.

\begin{table*}
\centering
\caption{Physical properties of the oils and the velocity $U$ of their moving contact lines.}
\begin{tabular}{ccccccc}
\hline
\multicolumn{1}{c}{Oil} & $\gamma_{\mathrm{oa}}$ (mN/m) &  $\gamma_{\mathrm{ow}}$ (mN/m) &  $\mu$ (mPa.s) &  $U_{\mathrm{theory}}$ (mm/s)  & $U_{\mathrm{exp}}^{\mathrm{av}}$ (mm/s)  &\\\hline
1-Octanol    & 27.5 & 8.4 & 7.4 &  47  & 30 &\\
1-Decanol    &  28.5& 8.6 &  12.2 &  29 & 25 &\\
1-Undecanol  & 26.6 & 8.8 & 17.5 & 21 & 18 &\\\hline
\end{tabular}
\label{tab:table-1}
\end{table*}

The water drop was deposited by a micro-pipette such that the encapsulated oil droplet was positioned near the three-phase contact line of water-air-solid. The footprint diameter of the water drop was 3--4 mm.
The water was slowly sucked out from the drop via a thin metallic needle of internal diameter 150 $\mu$m connected to a syringe operated by a syringe pump.
This led to slow receding velocities (3-70 $\mu$m/s) of the triple phase contact line of the water drop in the direction towards the encapsulated oil droplet, eventually leading to the coalescence of air-water-solid CL and oil-water-solid CL.

The dynamics exhibited upon the contact of the two triple phase contact lines was recorded  using a high-speed camera at 3000 fps from both bottom and side view.
The bottom views were recorded via an inverted microscope.
Additionally, this physical process was visualized using confocal microscopy. 
Nile-red (Sigma, $\lambda_{\mathrm{excit}} = 530$ nm) and fluorescein isothiocyanale-dextran (Sigma, $\lambda_{\mathrm{excit}} = 492$ nm) were used to selectively dye the oil and water phases, respectively.
The confocal images were recorded in resonant mode and bi-directional scanning was performed.
This limited the maximum scan rate to 30 frames per second (fps).  
Post-processing of the images acquired by the camera was performed using standard built-in functions of MATLAB R2017b (Canny edge-detection algorithm).

\section*{Author contributions}
B.D. and X.Z. initiated the research; H.Y., B.D. and P.K. performed the experiments; P.K. and H.Y. analyzed data; H.Y., P.K., D.L. and X.Z. contributed to the analytical framework and wrote the paper.

\section*{Ackownledgements}
H.Y. is grateful for valuable assistance from Dr Huanshu Tan in setting up the experiments . D.L acknowledges the European Research Council Advanced Grant-740479-DDD and X.Z. acknowledges the support of the Natural Sciences and Engineering Research Council of Canada (NSERC) and Future Energy Systems (Canada First Research Excellence Fund). 

%%%END OF MAIN TEXT%%%


\balance


%%%REFERENCES%%%
\bibliography{rsc} %You need to replace "rsc" on this line with the name of your .bib file
\bibliographystyle{rsc} %the RSC's .bst file

\end{document}